\documentstyle[twocolumn,aps]{revtex}

\begin{document}

\draft

\title{On linear coupling of acoustic and cyclotron waves in
plasma flows}

\author{Andria Rogava}
\address{Center for Plasma Astrophysics, Abastumani Astrophysical Observatory, Kazbegi str. $2^a$,
380060 Tbilisi, Georgia; and Abdus Salam International Centre for
Theoretical Physics, Trieste I-34014, Italy}

\author{Grigol Gogoberidze}
\address{Centre for Plasma Astrophysics, Abastumani Astrophysical Observatory, Kazbegi str. $2^a$,
380060 Tbilisi, Georgia}

\date{\today}

\maketitle

\begin{abstract}

It is found that in magnetized electrostatic plasma flows the
\emph{velocity shear} couples ion-acoustic waves with
ion-cyclotron waves and leads, under favorable conditions, to
their efficient reciprocal transformations. It is shown that in a
two-dimensional setup this coupling has a remarkable feature: it
is governed by equations that are {\it exactly} similar to the
ones describing coupling of \emph{sound waves} with \emph{internal
gravity waves} [Rogava \& Mahajan: Phys. Rev.~~E {\bf 55}, 1185
(1997)] in neutral fluid flows. Using another noteworthy quantum
mechanical analogy we calculate transformation coefficients and
give fully analytic, quantitative description of the coupling
efficiency for flows with low shearing rates.

\end{abstract}


\section{Introduction}

It is well-known that in flows with spatially inhomogeneous
velocities, due to the non-self-adjointness (``non-normality" [1])
of their linear dynamics, collective phenomena are strongly
influenced by the presence of the velocity shear [2]. In
non-uniformly flowing plasmas, for example, characteristic
nonperiodic modes
--- ``Langmuir vortices" [3], ``ion-acoustic vortices" [4] and
``dust-acoustic vortices" [5] ---  are arising due to the presence
of the inhomogeneous velocity field. They are analogous to so
called ``Kelvin modes" [6] (originally disclosed by Lord Kelvin
[7]) in plane Couette flows. Moreover, ``traditional" modes,
present in plasmas also in the absence of the flow, are also
interacting with ``ambient" flows and strongly modified by the
velocity shear. The modes may undergo transient (algebraic) growth
[6], and/or linear (adiabatic) amplification  [8] of their energy,
asymptotically persistent shear instabilities may also arise [9]
due to the kinematic complexity of ``parent" flows.

Another important and interesting issue, related with the
propagation of waves in  flows, is related with the ability of the
velocity shear to couple waves [10] and to ensure their mutual
transformations. Shear-induced wave couplings exist in
hydrodynamic systems (coupling of sound waves and internal gravity
waves [11]), in MHD (both electron-proton and electron-positron
plasmas), where velocity shear couples all three MHD wave modes
(Alfv\'en waves, slow magnetosonic waves and fast magnetosonic
waves) [12], and in dusty plasmas [5]. This phenomenon is
important in a wide variety of astrophysical applications,
including solar atmosphere and solar wind [13], pulsar
magnetosphere [14], galactic gaseous disks [15], self-gravitating,
differentially rotating dusty plasma clouds [16] and astrophysical
jets [17].

The mission of this paper is threefold. \emph{First}, it
demonstrates that the shear-induced transformations in plasmas are
{\it not} characteristic {\it only} to MHD systems but they exist
and are efficient also for low-frequency electrostatic waves: the
ion-sound waves (ISW) and the ion-cyclotron waves (ICW).
\emph{Second}, using recently developed method of the evaluation
of the transformation coefficients [18], we give full quantitative
and analytic description of the coupling efficiency for the case
of low shearing rate flows. \emph{Third}, we point out that the
mathematical description of this particular sort of wave coupling
is {\it exactly analogous} to the one in neutral fluid flows:
coupling of sound waves with internal gravity waves [11]! This
remarkable circumstance once more highlights the universality of
shear-induced phenomena in the physics of continuous media.

\section{Main Consideration}

Let us consider the simplest possible setup: the equilibrium
magnetic field ${\cal B}_0$ is homogeneous and directed along the
$X$ axis ${\vec {\cal B}}_0 \equiv ({\cal B}_0,0,0)$; the mean
velocity ${\vec {\cal U}}_0 \equiv (Ay,0,0)$ is also directed
along the $X$ axis and has a linear shear along the $Y$ axis. The
equilibrium densities of both electrons ${\cal N}_e$ and ions
${\cal N}_i$ are homogeneous. The temperature of electrons $T_e$
is constant, while ions are supposed to be cold $T_i=0$.

It is well known that ISW and ICW with phase velocities
considerably less than Alfv\'en speed ${\omega}/k_x{\ll}{\cal V}_A
\equiv {\cal B}_0/{\sqrt{4 {\pi}M{\cal N}_i}}$ ($M$ is the ion
mass) do {\it not} perturb the background magnetic field, and a
good approximation for the electric field [19] is: ${\bf
E}=-{\nabla}{\varphi}$.

Another common assumption for low-frequency electrostatic waves is
that the electrons ``thermalize along the field lines" and the
electron number density can be described by the Boltzmann
distribution [19]:
$$
{\cal N}_e={\cal
N}_0exp[e{\varphi}/T_e]{\approx}N_0[1+e{\varphi}/T_e]; \eqno(1)
$$
while dynamics of ions is governed by fluid equations of continuity and
motion:
$$
{\partial}_tN_i+{\nabla}{\cdot}(N_i{\bf V}_i)=0, \eqno(2)
$$
$$
[{\partial}_t+({\bf V}_i{\cdot}{\nabla})]{\bf V}_i={ e \over M }{\left
(-{\nabla}{\varphi}+{1 \over c}[{\bf V}_i{\times}{\bf B}]\right)}. \eqno(3)
$$

The basic system of linearized equations for ions, describing the evolution
of small-scale, 3D perturbations in this flow, takes the form:
$$
{\cal D}_tn_i+{\cal N}_0[{\partial}_xu_x+{\partial}_yu_y
+{\partial}_zu_z]=0, \eqno(4)
$$
$$
{\cal D}_tu_x+Au_y=-(e/M){\partial}_x{\varphi}, \eqno(5a)
$$
$$
{\cal D}_tu_y=-(e/M){\partial}_y{\varphi}+{\Omega}_cu_z, \eqno(5b)
$$
$$
{\cal D}_tu_z=-(e/M){\partial}_{z}{\varphi}-{\Omega}_cu_y, \eqno(5c)
$$
while the Poisson equation relates the perturbation of the
electric potential $\varphi$ with the number density perturbations
for electrons $n_e$ and ions $n_i$, respectively:
$$
{\left[{\partial}^2_x+{\partial}^2_y+{\partial}^2_z\right]}{\varphi}=
4{\pi}e(n_e-n_i). \eqno(6)
$$
Here ${\Omega}_c \equiv e{\cal B}_0/Mc
$ is the ion-cyclotron
frequency and ${\cal D}_t \equiv {\partial}_t+Ay{\partial}_x$ is
the convective derivative operator.

Employing for all perturbational variables appearing in the above
equations the ansatz $F(x,y,z;t)={\hat F} (t)exp[i(k_x x+k_y(t)
y+k_z z)]$ with $k_y(t) \equiv k_y(0)-Atk_x$. This ansatz
guarantees that ${\cal D}_tF = exp[i(k_x x+k_y(t) y+k_z
z)]{\partial}_t {\hat F}$ and we can reduce the initial set of
partial differential equations to the set of ordinary differential
equations for the amplitudes ${\hat F}$ (see for details, e.g.,
[1]). It is convenient to write these equations in dimensionless
notation. We define ${\omega}_c \equiv {\Omega}_c/C_{s}k_{x}$ as
the normalized ion-cyclotron frequency and ${\xi} \equiv
{\lambda}_Dk_{x}$, using for normalization conventional
definitions of the ion-sound speed $C_s \equiv (T_e/M)^{1/2}$ and
electron Debye length ${\lambda}_{D} \equiv (T_e/4{\pi}{\cal
N}_0e^2)^{1/2}$. Other notation are: $R \equiv A/C_{s}k_{x}$,
${\tau} \equiv C_{s}k_{x}t$, ${\beta}_0 \equiv k_{y}/k_{x}$,
${\beta}(\tau) \equiv {\beta}_0-R{\tau}$, ${\gamma}_0 \equiv
k_{z}/k_{x}$, $v_{x,y,z} \equiv {\hat u}_{x,y}/C_{s}$, $D \equiv
i(n_i/{\cal N}_0)$, $\Phi \equiv ie{\varphi}/MC_{s}^2$. In the
language of these terms the set of first order, ordinary
differential equations, derived from (4) and (5) takes the
following form:
$$
D^{(1)}=v_x+{\beta}(\tau)v_y+ \gamma
 v_z, \eqno(7)
$$
$$
v_x^{(1)}=-{\Phi}-Rv_y \eqno(8a)
$$
$$
v_y^{(1)}=-{\beta}(\tau){\Phi}+{\omega}_cv_z, \eqno(8b)
$$
$$
v_z^{(1)}=-{\gamma}{\Phi}-{\omega}_cv_y. \eqno(8c)
$$

Additionally we have the algebraic relation between $D$ and
$\Phi$, which follows from the Poisson equation (6):
$$
D={\left[1+{\xi}^2(1+{\beta}^2(\tau)+{\gamma}^2)\right]}F, \eqno(9)
$$

Note that ${\xi}{\omega}_c={\Omega}_c/{\Omega}_p={\cal V}_A/c$~~[$
{\Omega}_{p}{\equiv}(4{\pi}{\cal N}_0e^2/M)^{1/2}$ is the ion
plasma frequency]. When the Debye length is sufficiently small
(${\xi}{\ll}1$) the oscillations can be considered quasi-neutral
($D{\approx}F$). In the forthcoming consideration we shall employ
this approximation. However, we should bear in mind that due to
the ``$\textbf{k}(t)$-drift", induced by the existence of the
non-uniform motion, ${\beta}^2(\tau)$ sooner or later becomes
large enough and violates the quasineutrality condition.

As regards ${\omega}_c$, its value may be as less as greater than
unity, depending on the relative smallness of ${\xi}$ and ${\cal
V}_A/c$~~[$ {\Omega}_c/{\Omega}_p$] parameters.

The simplest sort of waves, which may be considered in this system
[19] are ones
propagating in the $X0Y$ plane (${\gamma}=0$). In
this case if we introduce an auxiliary notation
$Y{\equiv}-{\omega}_cD-{\beta} (\tau)v_z$, the system (7-8)
reduces to the following pair of coupled second-order differential
equations:
$$
Y^{(2)} + \omega_{1}^2Y + C_{12}(\tau)v_z= 0,
\eqno(10a)
$$
$$
v_z^{(2)} + \omega_{2}^2 (\tau) v_z + C_{21}(\tau)Y=0,
\eqno(10b)
$$
describing coupled oscillations with two degrees of freedom, with:
$C_{12}(\tau) = C_{21}(\tau) \equiv \beta(\tau)$ as the
\emph{coupling coefficients} and with $\omega_{1} \equiv 1$ and $
\omega_2^2(\tau) \equiv \omega_c^2+\beta^2(\tau)$  as the two
respective \emph{eigenfrequencies}. The presence of shear in the
flow ($R{\not=}0$) makes coefficients variable and opens the door
for mutual ICW--ISW transformations.

In the ``shearless" ($R=0$) limit (10) describe two independent
oscillations with frequencies:
$$
\Omega_{1,2}^2 = \frac{1}{2}\left[ \omega_1^2 + \omega_2^2 \pm
\sqrt{(\omega_1^2-\omega_2^2)^2+4C_{12}^2} \right]=
$$
$$
\nonumber\\
\frac{1}{2}\left[ \omega_c^2 + 1+\beta^2 \pm
\sqrt{(\beta^2+\omega_c^2+1)^2-4\omega_c^2} \right], \eqno(11)
$$
that can be readily identified as ISW and ICW frequencies
respectively. Corresponding eigenfunctions (sometimes called
\emph{normal variables}) are:
$$
\Psi_1=\frac{(\Omega_1^2-\omega_2^2)Y+C_{12}v_z}
{\sqrt{(\Omega_1^2-\omega_2^2)^2+C_{12}^2}}, \eqno(12a)
$$
$$
\Psi_2=\frac{(\Omega_1^2-\omega_2^2)v_z-C_{12}Y}
{\sqrt{(\Omega_1^2-\omega_2^2)^2+C_{12}^2}}. \eqno(12b)
$$

\subsection{Transformation coefficient}

The coupled oscillator systems similar to (10) with slowly varying
coefficients are well known in different branches of physics.
Mathematical methods for their analysis were first developed for
quantum mechanical problems: non-elastic atomic collisions [20]
and non-adiabatic transitions in two level quantum systems [21].
Later, similar asymptotic methods were successfully applied to
various other problems [22].

In [18] these efficient mathematical tools were used, for the
first time, for the study of the velocity shear induced coupling
and transformation of MHD waves. The problem which we are studying
now, also allows thorough probing by means of this asymptotic
method. We consider (most interesting for practical applications)
case $R\ll1$, when coefficient in (10) are slowly varying
functions of $\tau$ and, therefore, WKB approximation is valid
everywhere except nearby the \emph{turning points}
($\Omega_i(\tau_t)=0$) and \emph{resonant points }
($\Omega_1(\tau_r)=\Omega_2(\tau_r)$). Using (11) one can check
that the condition
$$
\Omega_i^{(1)} \ll {\Omega}_i^2. \eqno(13)
$$
is satisfied for both wave modes at any moment of time, or
equivalently, none of the turning points are located near the real
$\tau$-axis. From physical point of view this means, that there
are no (over)reflection phenomena [18] and only the \emph{resonant
coupling} between different waves modes with the same sign of

phase velocity can occur.

From (11) we also learn that there are two pairs of complex
conjugated resonant points of the first order\footnote{The
resonant point is said to be of the order $n$ if
$(\Omega_1^2-\Omega_2^2) \sim (\tau - \tau_{r})^{n/2}$ in the
neighborhood of the $\tau_{r}$.}:
$$
\beta(\tau_1^\pm) = \pm i(\omega_c -1),~~\beta(\tau_2^\pm) = \pm
i(\omega_c+1). \eqno(14)
$$
Therefore all the resonant points are located on the axis ${\rm
Re}[\beta(\tau)]=0$ in the complex $\tau$-plane and consequently,
the resonant coupling can take place only in a vicinity of the
point $\tau_\ast$ where $\beta(\tau_\ast)=0$. Generally, the time
scale of resonant coupling $\Delta \tau$ is of the order of
$\Delta \tau \sim R^{-1/2}$ [18] and the evolution of the waves is
adiabatic when
$$
|\beta(\tau)| \gg R^{1/2}. \eqno(15)
$$
If this condition is met the temporal evolution of the waves is
described by the standard WKB solutions:
$$
\Psi_i^\pm = \frac{D_i^\pm}{\sqrt {\Omega_i(\tau)}}e^{\pm i \int
\Omega_i(\tau)d\tau}, \eqno(16)
$$
where $D_i^\pm$ are WKB amplitudes of the wave modes with positive
and negative phase velocity along the $x$-axis, respectively. All
the physical quantities can be easily found by combining (12). One
can check that the energies of the involved wave modes satisfy the
standard adiabatic evolution condition:
$$
E_i = \Omega_i(\tau) (|{D_i^+}|^2 + |{D_i^-}|^2). \eqno(17)
$$

Let us assume that initially $\beta(0) \gg R^{1/2}$ and,
therefore, evolution of the waves is originally adiabatic. Due to
the linear drift in the ${\bf k}$-space, $\beta(\tau)$ decreases
and when the condition (15) fails to work, the mode dynamics
becomes non-adiabatic due to the resonant coupling between the
modes. The duration of the non-adiabatic evolution is given by
$\Delta \tau \sim R^{-1/2}$. Afterwards, when $\beta(\tau)
\ll-R^{-1/2}$, the evolution becomes adiabatic again. Denoting WKB
amplitudes of the wave modes before and after the coupling region
(i.e., for the $\tau<\tau_\ast-\Delta \tau/2$ and $\tau
 > \tau_\ast + \Delta \tau/2$) by ${D_{i,B}^\pm}$ and ${D_{i,A}^\pm}$
respectively and employing the formal analogy with the S-matrix of
the scattering theory [23] and the transition matrix from the
theory of multi-level quantum systems [24], one can connect
${D_{i,A}^\pm}$ with ${D_{i,B}^\pm}$ via the $4\times4$ transition
matrix:
$$
{\left(\matrix{{\bf D}^{+}_{A}\cr{\bf D}^-_{A} \cr}\right)}=
{\left(\matrix{{\bf T} & {\bf T}^{\pm} \cr {\bf T}^{\mp} & {\bf
T}^\ast \cr}\right)} {\left(\matrix{{\bf D}^{+}_{B}\cr{\bf
D}^-_{B} \cr}\right)}, \eqno(18)
$$
where ${\bf D}^{\pm}_{L}$ and ${\bf D}^{\pm}_{R}$; while
$1\times2$ matrices and ${\bf T}$, its Hermitian conjugated matrix
${\bf T}^\ast$, ${\bf T}^{\pm}$, and ${\bf T}^{\mp}$ are
$2\times2$ matrices.

None of the turning points are located near the real $\tau$-axis
and, therefore, only wave modes with the same sign of the phase
velocity along the $x$-axis can effectively interact. It is well
known [24] that in this case components of ${\bf T^{\pm}}$ and
${\bf T^{\mp}}$ are exponentially small with respect to the large
parameter $1/R$ and can be neglected. Consequently, (18)
decomposes and reduces to:
$$
{\bf D}^+_{A} = {\bf T} {\bf D}^{B}, \eqno(19a)
$$
$$
{\bf D}^-_{A} = {\bf T}^{\ast} {\bf D}^-_{B}. \eqno(19b)
$$

Since all coefficients in the (12) are real and $C_{12}=C_{21}$,
the matrix ${\bf T}$ is unitary [25], and
$$
\sum_j |T_{ij}|^2 =1. \eqno(20)
$$

Generally, this equation represents conservation of the wave
action. When $R \ll 1$ it transcribes  into the energy
conservation throughout the resonant coupling of wave modes [18].
The components of the transition matrix in (18) are complex, i.e.,
the coupling of different wave modes changes not only the absolute
values of $D_i^\pm $, but also their phases. The value of the
quantity $|T_{12}|^2$ represents a part of the energy transformed
during the resonant coupling of the modes. The absolute values of
the transition matrix components $|T_{12}|$ and $|T_{21}|$ are
called the \emph{transformation coefficients} of corresponding
wave modes. Unitarity of the ${\bf T}$ provides an important
symmetry property of the transition matrix:
$$
|T_{ij}|=|T_{ji}|, \eqno(21)
$$
i.e., transformation coefficients are reciprocally equal to each
other.

It is well known, that if in the neighborhood of the real
$\tau$-axis only a pair of complex conjugated first order resonant
points $\tau_{r}$ and $\tau_{r}^\ast$ exists, the transformation
coefficient is [18]:
$$
|T_{12}| = \exp\left( - \left| {\rm Im}
\int_{0}^{\tau_{r}}(\Omega_1-\Omega_2)d\tau \right| \right) [1+
O(R^{1/2})]. \eqno(22)
$$

As it was mentioned earlier, coupling between ICW and ISW can be
effective if $|\omega_c-1| \ll 1$. Otherwise transformation
coefficient is exponentially small with respect to the large
parameter $R^{-1}$ [24,25]. If this condition is satisfied the
resonant points $\tau_1^{\pm}$ [see (14)] tend to the real
$\tau$-axis and, hence, the effective coupling is possible.
Noting, that according to (13), in the neighborhood of the
resonant points:
$$
\Omega_1 - \Omega_2 \approx \sqrt{\beta^2(\tau)+(\omega_c-1)^2},
\eqno(23)
$$
and then  from (22) one can readily obtain:
$$
|T_{12}| \approx \exp \left[ - \pi (\omega_c-1)^2/4R \right].
\eqno(24)
$$

From the (14) we also see that if $\omega_c \rightarrow 1$, the
resonant points tend to the real $\tau$-axis. Then from (24) it
follows that $|T_{12}| \rightarrow 1$, i.e., one wave mode is
totally transformed into the another. On the other hand, if
$|\omega_c-1| \gg R^{1/2}$, transformation is negligible.

In the case of moderate or high shearing rates, similarly to the
case $R \ll 1$, one can show that the WKB approximation is valid
only when $\beta(\tau) \gg 1,R$. It means that the asymptotic
problem can still be formulated. However, when the shearing rate
is \emph{not} small, non-adiabatic evolution of the modes consists
of both \emph{transformation} and \emph{reflection} phenomena.
From mathematical point of view it means that, in general, all the
components of the transition matrix significantly differ from
zero. Conservation of the wave action remains valid and provides
following important relation [18]:
$$
\sum_j |T_{ij}|^2 - \sum_j |T_{ij}^{+-}|^2=1. \eqno(25)
$$

However, it should be also stressed that for high shearing rates
no analytical expressions for the components of the transition
matrix can be written explicitly.

\subsection{Hydrodynamical analogy}

The remarkable feature of our governing equations (10) is that
they are almost identical with the pair of equations from [11]
(numbered there as equations (16) and (17)):
$$
{\psi}^{(2)}+{\psi}+{\beta}(\tau)e=0, \eqno(26a)
$$
$$
e^{(2)}+{\left[W^2+{\beta}^2(\tau)\right]}e+
{\beta}(\tau){\psi}=0. \eqno(26b)
$$

These equations also describe coupling of two wave modes. But this
is totally different physical system: shear flow of a
gravitationally stratified neutral fluid, which sustains
\emph{sound waves} and \emph{internal gravity waves}. In [11] it
was shown that these modes are coupled through the agency of the
shear and may effectively transform into each other, providing the
condition $W{\simeq}1$ is met.

The presence of this analogy implies that all the details of the
transformation coefficient asymptotic analysis, which were given
above, can also be applied to the named hydrodynamic example of
the shear induced wave couplings. The only factual difference is
that we have to replace the dimensionless frequency of
electrostatic ion-cyclotron waves ${\omega}_c$ by the
characteristic dimensionless frequency $W$ of the internal gravity
waves.

This analogy allows also to predict that within the electrostatic
problem we should have a new kind of electrostatic {\it beat
waves} just as they exist in the hydrodynamical problem! Beat
waves are excited when an initial perturbation propagates almost
along the flow axis (${\beta}_0 {\ll}1$ and when, additionally,
${\tau}_{*}$ is sufficiently large in comparison with the {\it
beat period} ${\Omega}_b{\equiv}{\Omega}_{1}- {\Omega}_{2}$.

\section{Conclusion}

Summarizing main properties of the resonant coupling of ICW and
ISW for small shearing rates ($R\ll 1$) we can state that: (a)
Only the wave modes with the same sign of the phase velocity can
effectively interact - there are no reflection phenomena; (b)The
duration of the effective coupling of the modes is of the order of
$\Delta\tau \sim R^{-1/2}$, i.e., resonant coupling is slow
compared to the wave period $\tau_\Omega \sim 1$ but fast enough
compared to the adiabatic change of the system parameters $\tau_a
\sim R^{-1}$; (c) The total energy of the modes is conserved
during the resonant coupling - the transformed wave is generated
at the expense of the energy of the initial wave mode; (d) The
mode coupling process is symmetric - transformation coefficient of
one mode into another one equals the coefficient of the inverse
process. (e) The transformation coefficients are given by (22) and
(24).

None of these features remain valid for moderate and high shearing
rates. If $R \ll 1$ is not satisfied, there are no \emph{'long'}
and \emph{'shirt' }timescales in the problem and all the processes
have approximately the same characteristic timescale. Hence, the
coupling of the modes represents some mixture of transformation
and reflection precesses, that are accompanied by the energy
exchange between the waves and the background flow.

The discovered ICW--ISW transformations are likely to be important
in a number of applications. One possible example is the problem
of the ICW observed by low altitude satellites and ground based
magnetometers [26]. Observational surveys indicate that these
waves are correlated with the ICW observed in the equatorial
magnetosphere. However, theoretical studies of the ICW propagation
from the magnetosphere to the ground suggest that these waves can
not penetrate through the Buchsbaum resonance and can not reach
ionospheric layers of the atmosphere. Thus, one could expect that
the magnetospheric ICW shouldn't be correlated with the
ionospheric ICW, while observational evidence shows the
correlation. Recently Johnson and Cheng [26] reconsidered this
problem and found that strong {\it mode coupling} occurs near the
$He^{+}$ and $O^{+}$ resonance locations. They argued that this
coupling may help the equatorial ICW to penetrate to ionospheric
altitudes.

It seems reasonable to admit that the {\it velocity-shear-induced
ICW-ISW coupling} may provide yet another mode transformation
mechanism, which in conjunction with the one, found by Johnson and
Cheng, may account for the penetration of the ICW through the
Buchsbaum resonance. This may work in a quite similar way to the
scenario given in the [27] for the penetration of the fast
magnetosonic waves (FMW) from the chromosphere to the corona
through the solar transition region. The idea is that a fraction
of the FMW transforms into the Alfv\'en waves (AW). The latter go
through the transition region up to the solar corona, where they
again become transformed into the FMW. In this way `shear-induced
oscillations of solar MHD waves' [27], ensures substantial
transport of the FMW through the transition region. It seems
plausible to admit that the similar reciprocal `swinging' of ICW
and ISW may allow some fraction of the ICW to penetrate through
the Buchsbaum resonance and to reach the low ionospheric
altitudes.

Finally, the remarkable \emph{exact} analogy  of the ICW-ISW
coupling with the coupling of internal gravity waves and sound
waves in hydrodynamic flows [11] points out, once again, at the
universal character of the velocity shear induced phenomena in the
physics of fluids and plasmas.

\section{Acknowledgements}

Andria Rogava wishes to thank Abdus Salam International Centre for
Theoretical Physics for supporting him, in part, through a Senior
Associate Membership Award.

\end{document}